
\input phyzzx

%
%

\def\Roman#1{\uppercase\expandafter{\romannumeral #1}.}
\sectionstyle={\Roman}
\def\section#1{\par \ifnum\the\lastpenalty=30000\else
   \penalty-200\vskip\sectionskip \spacecheck\sectionminspace\fi
   \global\advance\sectionnumber by 1
   \xdef\sectionlabel{\the\sectionstyle\the\sectionnumber}
   \wlog{\string\section\space \sectionlabel}
   \TableOfContentEntry s\sectionlabel{#1}
   \noindent {\caps\uppercase\expandafter{\romannumeral\the\sectionnumber
.}\quad #1}\par
   \nobreak\vskip\headskip \penalty 30000 }

\def\subsection#1{\par
   \ifnum\the\lastpenalty=30000\else \penalty-100\smallskip \fi
   \noindent\enspace{#1}\enspace \vadjust{\penalty5000}}

\font\titlefont=cmcsc10 at 14pt
\font\namefont=cmr12
\font\placefont=cmsl12
\font\abstractfont=cmbx12

\font\sxrm=cmr6          \font\sxmi=cmmi6             
  \font\sxsy=cmsy6           
  
\font\nnrm=cmr9          \font\nnmi=cmmi9
  \font\nnsy=cmsy9           \font\nnex=cmex10 at 9pt
  \font\nnit=cmti9           \font\nnsl=cmsl9
  \font\nnbf=cmbx9
\def\nnpt{\def\rm{\fam0\nnrm}%
  \textfont0=\nnrm \scriptfont0=\sxrm \scriptscriptfont0=\fiverm
  \textfont1=\nnmi \scriptfont1=\sxmi \scriptscriptfont1=\fivei
  \textfont2=\nnsy \scriptfont2=\sxsy \scriptscriptfont2=\fivesy
  \textfont3=\nnex \scriptfont3=\nnex \scriptscriptfont3=\nnex
  \textfont\itfam=\nnit \def\it{\fam\itfam\nnit}%
  \textfont\slfam=\nnsl \def\sl{\fam\slfam\nnsl}%
  \textfont\bffam=\nnbf \def\bf{\fam\bffam\nnbf}%
  \normalbaselineskip=11pt
  \setbox\strutbox=\hbox{\vrule height8pt depth3pt width0pt}%
  \let\big=\nnbig \let\Big=\nnBig \let\bigg=\nnbigg \let\Bigg=\nnBigg
  \normalbaselines\rm}
\skip\footins=0.2in
\dimen\footins=4in
\catcode`\@=11
\def\vfootnote#1{\insert\footins\bgroup\nnpt
  \interlinepenalty=\interfootnotelinepenalty
  \splittopskip=\ht\strutbox
  \splitmaxdepth=\dp\strutbox \floatingpenalty=20000
  \leftskip=0pt \rightskip=0pt \spaceskip=0pt \xspaceskip=0pt
  \parfillskip=0pt plus 1fil
  \setbox1=\hbox{*}\parindent=\wd1\let\enspace=\null
  \hangafter1\hangindent\parindent\textindent{#1}\footstrut
  \futurelet\next\fo@t}
\catcode`\@=12
\def\footnoterule{\kern-3pt \hrule width2truein \kern 3.6pt}

\footline={\iftitlepage\hfil\global\titlepagefalse
\else\hss\tenrm\folio\hss\fi}

\newif\iftitlepage

\def\center{\parindent=0pt\leftskip=1in plus 1fill\rightskip=1in plus 1fill}

\def\preprint#1//#2//#3//{\baselineskip=14pt{\rightline{#1}\par}\vskip
.1in{\rightline{#2}\par}
{\rightline{#3}}\medskip}

\def\preprintb#1//#2//{\baselineskip=14pt{\leftline{#1}\par}{\leftline{#2}}}

\def\title#1\par{{\center\baselineskip=16pt
  \titlefont{#1}\par}}
\long\def\author #1// #2// #3// #4//{{\baselineskip=14pt\parskip=0pt
   \center\namefont{#1}\par\placefont#2\par#3\par#4\par}\par}

\long\def\abstract#1//{%
  \centerline{\abstractfont Abstract}\par
  {\baselineskip=14.5pt\advance\leftskip by 0pc\advance\rightskip by
0pc\parindent=10pt
  \def\enspace{\kern.3em}
  \noindent #1\par}}

\newcount\refno\refno=0
\long\def\rfrnc#1//{\advance\refno by 1
  $ $\llap{\hbox to\leftskip{\hfil\the\refno.\enspace}}#1\par\medskip}

\def\rl{\rightline}

\def\t1{{\tilde 1}}

\def\a3{\alpha_3(m_{Z^0})}

\newdimen\jbarht\jbarht=.2pt
\newdimen\vgap\vgap=1pt
\newcount\shiftfactor\shiftfactor=12

\catcode`\@=11
\def\jbarout{\setbox1=\vbox{\offinterlineskip
  \dimen@=\ht0 \multiply\dimen@\shiftfactor \divide\dimen@ 100
    \hsize\wd0 \advance\hsize\dimen@
  \hbox to\hsize{\hfil
    {\multiply\dimen@-2 \advance\dimen@\wd0
    \vrule height\jbarht width\dimen@ depth0pt}%
    \hskip\dimen@}%
  \vskip\vgap\box0\par}\box1}
\def\jbar#1{\mathchoice
  {\setbox0=\hbox{$\displaystyle{#1}$}\jbarout}%
  {\setbox0=\hbox{$\textstyle{#1}$}\jbarout}%
  {\setbox0=\hbox{$\scriptstyle{#1}$}\jbarout}%
  {\setbox0=\hbox{$\scriptscriptstyle{#1}$}\jbarout}}
\catcode`\@=12

\def\simlt{\mathop{\lower.4ex\hbox{$\buildrel<\over\sim$}}}
\def\simgt{\mathop{\lower.4ex\hbox{$\buildrel>\over\sim$}}}


\def\rg{\lower 2pt\hbox{$\buildrel > \over \sim$}}
\def\rl{\lower 2pt\hbox{$\buildrel < \over \sim$}}

\advance\vsize by 1truein 
\titlepagetrue

\preprint//{MIU-THP-94/68}//{June, 1994}//
\vskip 1truein
\title Using Gauge Coupling Unification and Proton Decay to Test Minimal
Supersymmetric SU(5)

\bigskip
\medskip
\author
John S. Hagelin, S. Kelley \break and Veronique Ziegler//
\vskip.1in
Department of Physics//
Maharishi International University//
Fairfield, IA  52557-1069, USA//

\bigskip
\medskip
\centerline {Abstract}

{\leftskip=.3truein\rightskip=\leftskip\baselineskip=16pt\noindent
We derive a one-loop expression, including all thresholds, for the mass of
the proton decay mediating color triplets, $M_{D^c}$, in minimal
supersymmetric SU(5). The result for $M_{D^c}$ does not depend on
other heavy thresholds or extra representations with SU(5) invariant
masses which might be added to the minimal model.
 We numerically correct our result to two-loop
accuracy.  Choosing inputs to maximize $M_{D^c}$ and $\tau_P$,
within experimental limits on the inputs and a $1~TeV$ naturalness
bound, we derive a strict bound $\alpha_3>0.117$.  We discuss how this bound
will change as experimental limits improve.
Measurements of $\alpha_3$ from deep inelastic scattering and the
charmonium spectrum are below the bound $\alpha_3>0.117$ by
more than $3\sigma$.  We briefly review several ideas of how to resolve the
discrepancy between these low values of $\alpha_3$ and the
determinations of $\alpha_3$ from LEP event shapes.\par}
\vskip 1.5in
\preprintb{MIU-THP-94/68}//{June, 1994}//

\endpage
\advance\vsize by -1truein 
\baselineskip=18pt

\pagenumber 1
A common criticism of unified theories is that physics at a
grand-unified scale of $10^{16}~GeV$ or a string-unified scale of
$10^{18}~GeV$ cannot be tested.  Indeed, a crucial ingredient of
unified models is the accurate reproduction of the highly
successful Standard Model predictions.  However, $\it{precision}$
measurements at low energies $\it{can}$ probe physics at much higher
energies.  Proton decay [1], gauge coupling unification [2],
flavor changing neutral currents [3], and sparticle spectroscopy [4]
all provide windows to physics at unified scales which can be
used to discriminate between different theories.
The minimal non-supersymmetric GUT has already been ruled out
by constraints from gauge coupling unification and proton
decay.  The purpose of this paper is to constrain the allowed
parameter space of minimal supersymmetric SU(5) [5] from the combined
experimental constraints on the low-energy gauge couplings and proton decay.
We
obtain a very general result that $\alpha_3>0.117$\footnote*{ Note that
$\alpha_e$, $\sin^2\theta$, and $\alpha_3$ are in the ${\jbar{MS}}$ scheme
renormalized at $m_Z$ throughout this paper, except where otherwise indicated.}
in the minimal model. This
result also holds for extended models containing the minimal model field
content plus extra representations with SU(5) invariant masses.

Although this limit is comfortably within the range of $\alpha_3$
extracted from LEP event shapes [6,7], measurements of $\alpha_3$ from
deep inelastic scattering [8], and the charmonium spectrum [9] fall
at least $3\sigma$ below this bound on $\alpha_3$.  This disagreement
between different types of measurements of $\alpha_3$ has not been
resolved, though several interesting proposals exist [10,11,12,13].  The
situation is similar to that in the electroweak sector, where different
types of measurements do not agree unless radiative corrections
are properly included.  However, radiative corrections to the
strong coupling are more difficult.  The definitive resolution
of conflicting measurements of $\alpha_3$ is an extremely important
project.  If the real value
of $\alpha_3$ lies close to that presently extracted from deep
inelastic scattering and the charmonium spectrum, we will have
ruled out yet another GUT.  If the LEP event shape values are correct,
supersymmetric SU(5) will survive even if proton decay is not
observed in the next round of experiments.

Minimal supersymmetric SU(5) is an amazingly complete model
which predicts the Standard Supersymmetric Model (SSM) as its
low-energy limit.  However, it is not a final model.
The fine-tuning problem could possibly be solved at the expense
of complications arising in non-minimal models.  A second
question is how to incorporate quantum gravity into the model.  General
considerations of quantum gravitational effects introduce
additional uncertainties into the calculations and blur predictions [14].
String theory could ultimately quantify the effects of quantum
gravity and resharpen the predictions.  However, there are
still obstacles to deriving useful predictions from the string, as well as
obtaining the scalar adjoint required to break SU(5) [15].

Considering that minimal supersymmetric SU(5) is most likely not
the complete theory of nature, why should one attempt to seriously
test it against experiment?  We feel that the most important reason
is to develop tools and techniques, which will be useful for more complete
models, in the context of a simple, definite model.  It
is difficult to develop techniques for testing string theories when
there are thousands to choose from, each with many parameters
beyond the present calculational ability to specify.  By taking
a specific, definite model like supersymmetric SU(5), which can
be rigorously compared to experiment, we can get some idea of what
must be done to test string models.

In the step approximation for thresholds, the
$SU(3)\times SU(2)\times U(1)$ couplings should unify at
a scale $M_X=max(M_V,M_{\Sigma},M_{D^c})$ where $V$ denotes the
superheavy vector superfields,
$\Sigma$ denotes the superheavy adjoint SU(3) and SU(2) remnants of the
adjoint higgs, and $D^c$ denotes the proton-decay mediating color triplets.
{}From this unification condition, the following predictions
result [16].
$$\eqalignno{
  ln\Bigl({M_X\over m_Z}\Bigr)&={\pi\over 6} \biggl({3\over 5\alpha_e}-{8\over
   5\alpha_3}\biggr)+{\sum\atop l} C_m(l)ln\Bigl({m_l\over
m_Z}\Bigr)-{\sum\atop
   h} C_m(h) ln\Bigl({M_X\over M_h}\Bigr)+\delta_m\cr
& &(1.a)\cr
& &\cr
\sin^2\theta&=0.2+{7\alpha_e\over
15 \alpha_3}+{\alpha_e\over 20\pi}\biggl[{\sum\atop l}
C_s(l) ln\Bigl({m_l\over m_Z}\Bigr)-{\sum\atop h} C_s(h)ln\Bigl({M_X\over
M_h}\Bigr)\biggr] +\delta_s\cr
&  &(1.b)\cr
& &\cr}$$
$${1\over \alpha_X}={1\over4}\biggl({3\over
5 \alpha_e}+{12\over 5\alpha_3}\biggr)+{1\over 20\pi}\biggl[{\sum\atop l}
C_a(l)ln\Bigl({m_l\over m_Z}\Bigr)-{\sum\atop h} C_a(h)ln\Bigl({M_X\over
M_h}\biggr)\Bigr]
                                     +\delta_a\eqno{(1.c)} $$
where the sums over the light $m_l\rg m_Z$, and heavy $M_h\rl M_X$, thresholds,
$$\eqalignno{
  C_m&={b_y\over 12}+{b_2\over 20}-{2\over 15}b_3&(2.a)\cr
  C_s&= {10\over 3}b_y-8 b_2 +{14\over 3}b_3&(2.b)\cr
  C_a&= {5\over 2}b_y +{3\over 2} b_2 +6 b_3&(2.c)\cr
}$$
and the correction terms $\delta_i=\delta_i(gauge)+\delta_i(Yukawa)$
include two-loop gauge and Yukawa effects.  Choosing
the $\delta_i(gauge)$ linear in $\alpha_3$ so that the analytic
approximations, Equations (1.a,b,c), match numerical two-loop calculations
at $\alpha_3=0.11$ and $\alpha_3=0.13$ gives
$$\eqalignno{
  \delta_m(gauge)&=-0.0868852 - 1.1444784\alpha_3&(3.a)\cr
  \delta_s(gauge)&=0.00127231 +0.0147977\alpha_3&(3.b)\cr
  \delta_a(gauge)&=-0.2385283 - 3.6119007\alpha_3&(3.c)\cr
}$$
Figure 1 shows that this works extremely well for experimentally
interesting values of $\alpha_3$.

 Equations (1.a) and (1.b) can be used to solve for the mass
of the proton-decay mediating triplet [17]:
$$\eqalignno{
ln\Bigl({M_{D^c}\over m_Z}\Bigr)&= {3\pi\over\alpha_e}\Bigl({0.2324 \pm{0.0003}
- 1.03\times{10^7}[m_t^2-(138 GeV)^2] - {5\alpha_e\over{9\alpha_3}} -{1\over
6}}\Bigr)\cr
\noalign{\smallskip}
&+{\sum\atop l} C_d(l)ln\Bigl({m_l\over
m_z} \Bigr)-{\sum\atop {h-D^c}} C_d(h) ln\Bigl({M_X\over M_h}\Bigr)+\delta_d
&(4)
}$$
where
$$\eqalignno{
   C_d&= C_m - {3\over20}C_s= {-5\over12}b_y + {5\over4}b_2 -
{5\over6}b_3&(5.a)\cr
   \delta_d&= -{3\pi\over\alpha_e}\delta_s + \delta_m&(5.b)\cr
}$$
and we have used the experimental value of $\sin^2\theta$
with explicit top mass dependence [18].

Notice that $C_d$ vanishes for the representation content of $M_V$ and
$M_\Sigma$ [17].  This means that $M_{D^c}$ only depends on the low-energy
parameters in the model and not on the other heavy
scales.  Additional representations beyond the minimal model
with SU(5) invariant masses will not change this result.
Therefore, in Eq. (4) $${\sum\atop {h-D^c}} C_d(h) ln\Bigl({M_X\over M_h}\Bigr)
=
0.\eqno{(6)}$$ Writing out the contribution from light thresholds to Eq. (4)
gives
$${\sum\atop {l}} C_d(l) \ln\Bigl({m_l\over m_Z}\Bigr) = {11\over
24}\ln\Bigl({m_t\over {m_Z}}\Bigr)
+ {5\over 3}\ln\Bigl({m_{\tilde w}\over {m_{\tilde g}}}\Bigr) + {1\over
6}\ln\Bigl({m_h\over
{m_Z}}\Bigr) +{2\over 3}\ln\Bigl({\mu\over {m_Z}}\Bigr)
+ g_d(y,w)\eqno{(7)},$$ where $g_d(y,w)$ includes
the threshold effects of squarks and sleptons [19].

To derive the most conservative bound, we must
maximize Eq. (4) within the experimental limits of the inputs.
Calculations of Yukawa effects on gauge coupling unification
give [18]
$$\delta_s(Yukawa)>-0.0004,\quad \delta_m(Yukawa)<0.06\eqno(8)$$

\noindent Taking into consideration experimental limits and using a
$1~TeV$ naturalness bound, $M_{D^c}$ is maximized by taking
$$\sin^2\theta=.2327-1.03\times 10^{-7}(m_t^2-138^2~GeV^2)\eqno{(9.a)}$$
$$\alpha_e={1\over127.8}\eqno{(9.b)}$$
$$m_t=130~GeV\eqno{(9.c)}$$
$$ m_h=\mu=1~TeV\eqno{(9.d)}$$
$$g(y,w)=0\footnote*{Because $C_m$, $C_s$, and $C_d$ vanish
for complete $SU(5)$ representation, $g_m$, $g_s$, and $g_d$,
which include the threshold corrections for the squarks and sleptons,
are close to zero over the whole parameter space. Under simple assumptions
for the stop mass matrix, $\max[g_d(y,w)]=0$.}\eqno{(9.e)}$$

The threshold effects of the gauginos require more careful
consideration.
 Since
$$m_{\tilde i}=m_{1/2}{{\alpha_i(m_{\tilde i})}\over{\alpha_X}}\eqno{(10)}$$
we have
$${m_{\tilde w}\over{ m_{\tilde g}}}={{\alpha_2(m_{\tilde w})}\over
{\alpha_3(m_{\tilde g})}}\eqno{(11)}$$
Below the wino threshold, $b_2$ is negative and $\alpha_2$
decreases. Therefore, $\alpha_2$ at the wino mass is bounded by
$$\alpha_2(m_{\tilde w})<\alpha_2(m_Z).\eqno{(12)}$$
Below the gluino threshold, $b_3>-5$. Therefore $\alpha_3$ at a gluino mass
less
than $1~TeV$ is bounded by
$$\alpha_3(m_{\tilde g})>
{\alpha_3(m_Z)\over {1+{5\over {2\pi}}\alpha_3(m_Z)\ln\Bigl({1~TeV\over
{m_Z}}\Bigr)}}.\eqno{(13)}$$
This results in a bound
$${m_{\tilde w}\over m_{\tilde g}}
<{\alpha_e\over{{\sin}^2\theta~\alpha_3}}\Bigl[
{1 + {5\over {2\pi}}\alpha_3\ln{\Bigl({1~TeV\over {m_Z}}\Bigr)}\Bigr]} =
{0.0336\over {\alpha_3}} + 0.0641\eqno{(14)}$$
Finally, the resulting bound on $M_{D^c}$ as a function of the strong
coupling is
$$\ln\Bigl({{M_{D^c}\over {m_Z}}}\Bigr)<80.8831 - {5\pi\over {3\alpha_3}}
- 18.9705\alpha_3
+ {5\over 3}\ln\Bigl[{{0.0336\over {\alpha_3}}+0.0641}\Bigr]\eqno{(15)}$$

Turning to proton decay,
a detailed exploration of proton decay reveals
that with a $1~TeV$ naturalness bound, the most conservative
bound on $M_{D^c}$ is given by [17]
$$ln\Bigl({M_{D^c}\over m_Z}\Bigr)>31.7\eqno{(16)}$$

\noindent This result uses a bound of $0.0003<\beta/GeV^3<0.003$ on the
hadronic matrix element.
We will use a value of the hadronic matrix element from lattice calculations
$\beta=(5.6\pm 0.5)\times 10^{-3}~GeV^3$ [20] which gives a bound on $M_{D^c}$.
$$ln\Bigl({M_{D^c}\over m_Z}\Bigr)>32.2\eqno{(17)}$$

\noindent Figure 2 plots the two bounds (15) and (17) in the $M_{D^c},\alpha_3$
plane revealing
that minimal supersymmetric SU(5) is compatible with proton
decay and coupling constant only if $\alpha_3>0.117$.  This analytic result
matches
well with  $\alpha_3>0.118$ from a similar numerical calculation over
large regions of the SU(5) parameter space [21].

One would not
expect our analytic bound to be saturated in a full
numerical calculation for any point in SU(5) parameter space. This
is because our result uses an upper bound
on $M_{D^c}$ from coupling constant unification and a lower bound on $M_{D^c}$
from proton decay which are independent, and each bound
has been extremized for each input separately.  Because of
correlations in the spectrum, neither bound is likely
to be saturated nor are the upper and lower bounds likely
to be simultaneously realized.  Our analytic bound is thus
very conservative and can be trusted over the whole parameter
space of the model without questioning the intricacies of a
computerized search over a large dimension parameter space.

It is interesting to see the effect of improvements in the
various experimental bounds on the minimum value of $\alpha_3$
compatible with minimal supersymmetric SU(5).  A factor of
ten improvement in the bounds on the proton lifetime would
give $\alpha_3>0.120$, while a top quark mass of $m_t\ge 175~GeV$ would give
$\alpha_3>0.121$.  Both a tenfold improvement in the proton
lifetime bound and a top mass $m_t\ge 175~GeV$ would give $\alpha_3>0.125$.
Since the bound on $M_{D^c}$ from proton decay is proportional to $m_{\tilde
w}/m^2_{\tilde q,\tilde
l}$, an increase in the upper bound on the wino mass or evidence
that the squarks and sleptons are below $1~TeV$ would  result
in a more stringent bound on $\alpha_3$.  A factor of four increase
in $m_{\tilde w}/m^2_{\tilde q,\tilde l}$ would give
$\alpha_3>0.121$.  Conversely a factor of four decrease in
$m_{\tilde w}/m^2_{\tilde q,\tilde l}$ due to
an increase in the naturalness bound
from $1~TeV$ to $2~TeV$ would give $\alpha_3>0.113$.

{}From an experimental standpoint, even the bound $\alpha_3>0.125$ is
compatible with present extractions of
$\alpha_3$ from LEP event shapes. Values of $\alpha_3 = 0.120\pm 0.006$
from OPAL [6] and $\alpha_3= 0.123\pm 0.006$ from DELPHI [7] represent
typical central values and error bars,
although the central values are sensitive to what observable is used
and the details of the analysis method.
However, there are indications that
the strong coupling might be much smaller.  Analysis of deep
ineleasic scattering data gives
$\alpha_3=0.108\pm 0.002$ [8].  Lattice gauge theory
calculations based on the 1P-1S splitting in charmonium give
$\alpha_3=0.105\pm 0.004$ [9].  Resolving the discrepency between the
high energy measurements of $\alpha_3$ at LEP and various low energy
measurements has been an important issue in GUTS for some time [22].

Tree level values of couplings from different experiments
should not match precisely; radiative corrections must be included.
The precision of different measurements of electroweak parameters
has reached the sensitivity where theory is being
rigourosly tested at the loop level, and radiative corrections must
be properly included to see the consistancy of different types of
measurements [23].  It is likely that there is a similar, though more
computationally intensive, explanation of the discrepencies
between different measurements of $\alpha_3$.
Several ideas regarding the details of this explanation
already exist.  A proposed method for summing the
higher-order corrections to LEP event shapes reduce the LEP event shape
extractions of $\alpha_3$ by about 0.012 [10].  Attempts to systematically
resolve the scale and scheme ambiguities [11] in determining
$\alpha_3$ have resulted in a value of $\alpha_3=0.107\pm 0.003$
extracted from LEP event shapes [12].
 A primary motivation for the light gluino is the reconciliation
of different measurements of $\alpha_3$ due to the modification of the
running of $\alpha_3$ in the presence of light gluinos [13]. With light
gluinos, deep inelastic scattering measurements give $\alpha_3=0.124\pm0.001$
[8].

However, in the context of radiative electroweak breaking in the SSM,
which minimal supersymmetris SU(5) reduces to after integrating out the
superheavies [24], the light gluino scenario is in trouble!
Simple considerations of the chargino and neutralino sector severely restrict
the parameter space [25].
Moreover, explicit calculations of radiative electroweak breaking show
that the small area of parameter space compatible with other experimental
limits
has a top mass of about $114~GeV$ [26], which
is well below present experimental limits [27].
Nonuniversal supersymmetry breaking may be able to save the light gluino
scenario. However, because of the large ratio of $m_{\tilde q}/m_{\tilde g}$,
the light gluino scenario gives a relic density hundreds of times larger than
compatible with the age of the universe.
Introducing R parity breaking into the model would probably be required to
remedy this dark matter problem.

In conclusion, we have deduced a conservative bound, $\alpha_3 > 0.117$,
in the minimal supersymmetric SU(5) model. Confirmation of a high top mass and
improvements in proton lifetime bounds would substantially increase this bound.
Although compatible with values of $\alpha_3$ extracted from LEP event shapes,
the bound is in contradiction with measurements of $\alpha_3$ from deep
inelastic
scattering [8] and the charmonium spectrum [9].
Resolution of the discrepancies between different measurements of
the strong coupling is of extreme importance for testing GUT scale
physics.
Although proper incorporation of radiative corrections to the measurements
of $\alpha_3$ will most likely remove all discrepancies, there is no concensus
on the details. The light gluino scenario brings most measurements of
$\alpha_3$
into agreement with those from LEP event shapes.
However, the light gluino scenario has severe phenomenological problems.

Unless the gluino is light or another source of radiative correction
substantially
increases the value of $\alpha_3$ extrated from deep inelastic scattering
and the charmonium spectrum, the bound $\alpha_3 > 0.117$ may signal the
demise of minimal supersymmetric SU(5).
\bigskip
{\bf Acknowledgements: }We thank H. Pois and B. Wright for useful discussions.
The work of John Hagelin is supported in part by National Science Foundation
under grant No. PHY-9118320.
\vfill
\eject
\eject
\centerline{\bf REFERENCES}
\bigskip
\item{[1]  }H.~Georgi and S.L.~Glashow, Phys. Rev. Lett. {\bf 32}(1974) 438.

\item{[2]  }H.~Georgi, H.~Quinn and S.~Weinberg, Phys. Rev. Lett.  {\bf 33}
(1974) 451.

\item{[3]  }H.~Georgi, Phys. Lett. {\bf 169B} (1986) 231;\hfill\break
            L.J.Hall, V.A.~Kostelechy and S.~Raby, Nucl. Phys. {\bf  B 267}
(1986) 415.

\item{[4]  }A.E.~Faraggi, J.S.~Hagelin, S.~Kelley and D.V.~Nanopoulos, Phys.
Rev. {\bf D 45}
(1991) 3272.

\item{[5]  }S.~Dimopoulos and H.~Georgi, Nucl. Phys. {\bf B 193} (1981)
150;\hfill\break
           N.~Sakai, Z. Phys. {\bf C 11} (1982) 153;\hfill\break
            A.~Chamseddine, R.~Arnowitt and P.Nath, Phys. Rev. Lett. {\bf 105}
(1982) 150.

\item{[6]  }OPAL collaboration, Z. Phys. {\bf C 59} (1993) 1.

\item{[7]  }DELPHI collaboration, Z. Phys. {\bf C 59} (1993) 21.

\item{[8]  }J.~Blumlein, J.~Botts, Phys. Lett. {\bf B 325} (1994) 190.

\item{[9]  }A.X.~Khadra, G.~Hockney, A.S.~Kronfeld and P.B.~Mackenzie, Phys.
Rev. Lett. {\bf 69} (1992)
729.

\item{[10]  }J.~Ellis, D.V.~Nanopoulos and D.A.~Ross, Phys. Lett. {\bf B267}
(1991) 132.

\item{[11]  }S.J.~Brodsky and H.J.~Lu, Preprint SLAC-PUB 6389 (1993).

\item{[12]  }G.~Kramer and B.~Lampe, Z. Phys. {\bf 339} (1991) 189.

\item{[13]  }J.~Ellis, D.V.~Nanopoulos and D.A.~Ross, Phys. Lett. {\bf B 305}
(1993) 375.

\item{[14]  }L.J.~Hall and U.~Sarid, Phys. Rev. Lett. {\bf 70} (1993) 2673.

\item{[15]  }H.~Dreinel, J.L.~Lopez, D.V.~Nanopoulos and D.~Reisa, Phys.
Lett.\break {\bf B216}
(1989) 289.

\item{[16]  }J.~Ellis, D.V.~Nanopoulos and S.~Kelley, Phys. Lett. {\bf B 260}
(1991) 131.

\item{[17]  }J.~Hisano, H.~Murayama and J.~Yanagida, Nucl.Phys. {\bf B 402}
(1993) 46.

\item{[18]  }P.~Langacker and N.~Polonsky, Phys. Rev. {\bf D 47} (1993) 4028.

\item{[19]  }J.~Ellis, D.V.~Nanopoulos and S.~Kelley, Nucl. Phys. {\bf B 373}
(1992) 55.

\item{[20]  }M.B.~Gavela et al., Nucl Phys. {\bf B312} (1989) 269.

\item{[21]  }B.~Wright, Preprint MAD/PH/812 (1994).

\item{[22]  }J.S.~Hagelin and S.~Kelley, Preprint MIU-THP-92/61 (1992).

\item{[23] }G.~Altarelli, Preprint CERN-TH.7072/93 (1993).

\item{[24]  }J.L.~Lopez, D.V.~Nanopoulos, H.~Pois, S.~Kelley and K.~Yuang,
Nucl. Phys. {\bf B 98}
(1993) 3.

\item{[25]  }L.~Clavelli, P.~Coutter and K.~Yuang, Phys. Rev. {\bf D 47} (1993)
1973.

\item{[26]  }J.L.~Lopez, D.V.~Nanopoulos and X.~Wang, Phys. Lett. {\bf 313}
(1993) 241.

\item{[27]  }CDF Collaboration, Preprint FERMILAB-PUB-94-097-E (1994).

\bigskip \bigskip \bigskip
\bigskip \bigskip \bigskip

{\centerline{\bf Figure Captions}}
\bigskip
\noindent (1) Percentage differences between the fitted analytic results and
the explicit numerical
calculations for $\sin^2\theta$, $\ln\bigl({M_X\over m_Z}\bigr)$ and
$\alpha_X$.
The fit is chosen so that the difference vanishes at $\alpha_3=0.11$ and
$\alpha_3=0.13$.

\noindent (2) The allowed region in the $M_{D^c}$, $\alpha_3$ plane for the
minimal
supersymmetric SU(5) model.
The upper bound on $M_{D^c}$ is from gauge coupling unification, and
the lower bound on $M_{D^c}$ is from proton decay.
The minimal supersymmetric SU(5) model is only compatible with values of
$\alpha_3$
greater than 0.117.

\end